\documentclass[showpacs,preprintnumbers,twocolumn,prl,aps]{revtex4}%
\usepackage{graphicx}
\usepackage{dcolumn}
\usepackage{bm}
\usepackage{amssymb,amsmath}
\usepackage{amsmath}
\usepackage{amsfonts}
\usepackage{amssymb}%

\begin{document}
\preprint{To appear in Physical Review Letters}
\title{Observation of Kinetic Plasma Jets}
\author{S. K. P. Tripathi}
\altaffiliation{Present address: Physics and Astronomy, 
University of California, Los Angeles, Los Angeles, California 90095}
\author{P. M. Bellan}
\author{G. S. Yun}
\affiliation{Applied Physics, California Institute of Technology, Pasadena, California 91125}
\date{\today }

\begin{abstract}
Under certain conditions an intense kinetic plasma jet is observed to emerge
from the apex of laboratory simulations of coronal plasma loops. Analytic and
numerical models show that these jets result from a particle orbit instability
in a helical magnetic field whereby magnetic forces radially eject rather than
confine ions with sufficiently large counter-current axial velocity.
\end{abstract}
\maketitle
\begin{figure}
\includegraphics{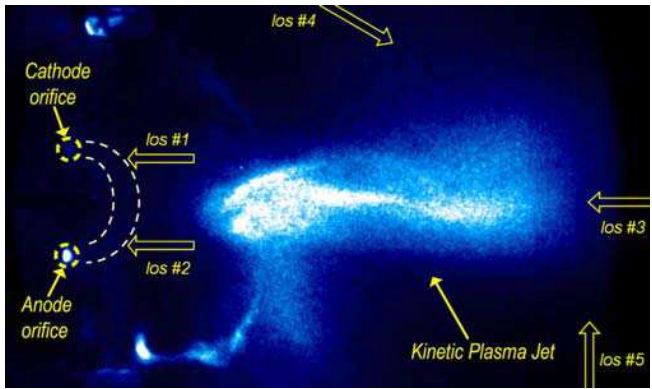}
\caption{(Color online) Kinetic plasma jet emanating from an argon 
laboratory loop. Dashed arch corresponds to initial plasma loop 
at $t\approx1.0$ $\mu$s as in Fig.\ \ref{fig:ar}(b). 
Arrows represent lines of sight (los\#) used for
spectroscopy; arrow widths represent $\sim 6$ mm
diameter of lines of sight. }%
\label{fig:jet}%
\end{figure}

%PACS, the Physics and Astronomy
%Classification Scheme.
%52.72.+v Laboratory studies of space- and astrophysical-plasma processes
%96.60.Pb Corona; coronal loops, streamers, and holes
%52.55.Ip Spheromaks
%52.55.Fa Tokamaks, spherical tokamaks
%\keywords{Suggested keywords}%Use showkeys class option if keyword display desired
%%%%%%%%%%%%% SECTION 1 %%%%%%%%%%%%%%%%%%%%%%%%%%%%%%%
%\section{Introduction\label{sec:intro}}
Many lab and space plasmas (e.g., solar coronal loops \cite{Tandberg},
spheromaks \cite{Bellan_book2000}, tokamaks \cite{Tokamak}, and magnetic
clouds \cite{Burlaga}) are presumed to be magnetic flux tubes filled with
plasma confined via magnetohydrodynamic (MHD) forces. However, confinement can
be significantly degraded in ways not predicted by MHD; e.g., in tokamaks,
ions resulting from neutral beams injected against the toroidal current
direction (counter-injection) exhibit severe orbit losses compared to
co-injection \cite{Mikkelsen,Egedal,McClements}. Related confinement
degradation may be the cause of small solar corona jets (e.g.,
surges)\ associated with canceling magnetic features \cite{Chae,Liu,Harrison}
and of coronal streamers emanating from magnetic neutral lines \cite{Li}. This
Letter reports that in certain circumstances an ion injected along the axis of
a magnetic flux tube will be magnetically \textit{ejected }from the flux tube
instead of being magnetically confined, i.e., the ion will have radially
unstable motion (RUM). This instability explains the severe orbit losses of
counter-injected ions in tokamaks and is likely relevant to similar situations
occurring in the solar corona. The instability, labeled as `Kinetic Plasma
Jet' in Fig.\ \ref{fig:jet}, was discovered experimentally and then modeled.
%%%%%%%%%%%%%%%%%%%%%%%%%%%%%%%%%%%%%%%%%%
%-----FIGURE 1 (Typical Jet)
%%%%%%%%%%%%%%%%%%%%%%%%%%%%%%%%%%%%%%%%%%

\begin{figure}
\includegraphics{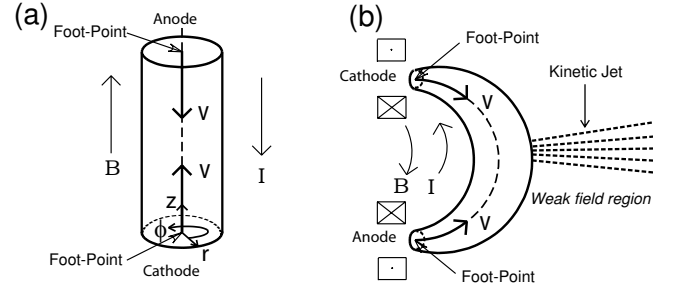}
\caption{(a) Flux tube geometry used in
the model. (b) Experimental configuration; $B$ outside the flux loop is
weakest on the right hand side (indicated as weak field region).}%
\label{fig:coordinate}%
\end{figure}

We first outline the physical basis for RUM. Consider a particle injected with
velocity $v_{z0}$ near the axis of a cylindrical flux tube having helical
magnetic field $\mathbf{B=}B_{\phi}\hat{\phi}+B_{z}\hat{z}.$ The flux tube
geometry is sketched in Fig.\ \ref{fig:coordinate}(a) and corresponds
to a straightened-out model of Fig.\ \ref{fig:coordinate}(b), our
laboratory configuration simulating a coronal loop \cite{Hansen04}. The $r$
component of $md\mathbf{v}/dt=q\mathbf{v\times B}$ is $m\ddot{r}-mr\dot{\phi
}^{2}=qr\dot{\phi}B_{z}-qv_{z}B_{\phi}$. If $v_{z}B_{\phi}=0$, then radial
force balance $\ddot{r}=0$ gives $\dot{\phi}=-qB_{z}/m=-\omega_{c}$, i.e., the
conventional cyclotron orbit. However, if $v_{z}B_{\phi
}\neq0$, then radial force-balance $\ddot{r}=0$ requires $\dot{\phi}^{2}%
+\dot{\phi}qB_{z}/m-v_{z}qB_{\phi}/mr=0$ so no real $\dot{\phi}$ solutions
exist if
\begin{equation}
B_{z}^{2}+4mv_{z}B_{\phi}/qr<0. \label{equation of motion criteria}%
\end{equation}
Thus, large negative $mv_{z}B_{\phi}/qr$ causes the radially \textit{outward}
force$\ -qv_{z}B_{\phi}$ to overwhelm the radially \textit{inward} force
$qr\dot{\phi}B_{z}$ so centrifugal force $mr\dot{\phi}^{2}$ is unbalanced.

%%%%%%%%%%%%%%%%%%%%%%%%%%%%%%%%%%%%%%%%%%
%-----FIGURE 2 (Co-ordinate system)
\begin{figure}
\includegraphics{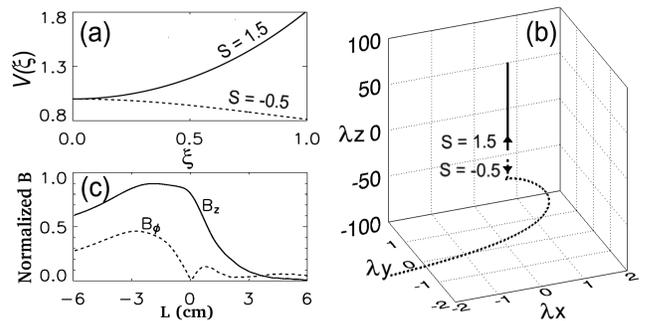}
\caption{(a)\ $S=1.5$ gives `valley'
(i.e., stable) effective potential $V(\xi)$ while $S=-0.5$ gives $\ $ `hill'
(i.e., unstable)$.$ (b) Numerically calculated particle trajectories. (c)
Probe measurement of flux tube magnetic field showing that field is weak on
right side as sketched in Fig.\ \ref{fig:coordinate}(b).}%
\label{fig:potential}%
\end{figure}

%%%%%%%%%%%%%%%%%%%%%%%%%%%%%%%%%%%%%%%%%%
We next use Hamiltonian arguments to show that satisfying
Eq.\ (\ref{equation of motion criteria}) leads to the particle being radially
ejected from the flux tube, i.e., RUM. In order to model the simplest
nontrivial situation, both $B_{z}$ and the axial current density $J_{z}$ are
assumed uniform within the flux tube so in the flux tube the vector potential
is $\mathbf{A}(r)=\hat{\phi}B_{z}r/2-\hat{z}\mu_{0}J_{z}r^{2}/4,$ the axial
flux is $\Phi=B_{z}\pi r^{2},$ and the axial current is $I=J_{z}\pi r^{2}.$
Using the Lagrangian $L=m(v_{r}^{2}+r^{2}\dot{\phi}^{2}+v_{z}^{2}%
)/2+qr\dot{\phi}A_{\phi}+qv_{z}A_{z}$, the canonical momenta $P_{\phi
}=\partial L/\partial\dot{\phi}$ and $P_{z}=\partial L/\partial v_{z}$ are
\begin{equation}
P_{\phi}=\ mr^{2}\dot{\phi}+qr^{2}B_{z}/2,\text{ \ }P_{z}=mv_{z}-\mu_{0}%
qJ_{z}r^{2}/4. \label{P-definition}%
\end{equation}
Because $\phi$ and $z$ are ignorable, both $P_{\phi}$ and $P_{z}$ are
invariants. A particle injected with velocity $v_{z0}$ along the the flux tube
axis (i.e., at $r=0)$ thus has the invariants%
\begin{equation}
P_{\phi}=0\text{, \ \ }P_{z}=mv_{z0}. \label{P-actual}%
\end{equation}
Combining Eqs.\ (\ref{P-definition}) and (\ref{P-actual}) gives $\dot{\phi
}=-\omega_{c}/2$ and $v_{z}=v_{z0}+\omega_{c}r^{2}\lambda/4$ where $\ $
$\lambda=\mu_{0}J_{z}/B_{z}=\mu_{0}I/\Phi$ is related to twist. The
Hamiltonian $H=m(v_{r}^{2}+r^{2}\dot{\phi}^{2}+v_{z}^{2})/2$ can be expressed
as $H=mv_{r}^{2}/2+f(r)$ where
\begin{equation}
f(r)=\frac{m\omega_{c}^{2}}{2\lambda^{2}}\left[  \frac{\lambda^{2}r^{2}}%
{4}+\left(  \frac{\lambda v_{z0}}{\omega_{c}}+\frac{\lambda^{2}r^{2}}%
{4}\right)  ^{2}\right]  \ \label{f}%
\end{equation}
is an effective potential. On defining $\xi=\left\vert \lambda\right\vert r,$
the dimensionless effective potential $V(\xi)=2\lambda^{2}f(r)/m\omega_{c}%
^{2}$ can be written as%
\begin{equation}
V(\xi)=\xi^{2}/4+\left(  \lambda v_{z0}/\omega_{c}+\xi^{2}/4\right)  ^{2}.
\label{V}%
\end{equation}
Equation (\ref{V}) gives $\partial V/\partial\xi=\xi/2+\lambda v_{z0}%
\xi/\omega_{c}+\xi^{3}/4$. Near the flux tube axis the $\xi^{3}/4$ term is
negligible, so negative $\partial V/\partial\xi$ near the axis corresponds to
having
\begin{equation}
S=\lambda v_{z0}/\omega_{c}+0.5<0. \label{S}%
\end{equation}
Our main result is that if $S<0$ so $\partial V/\partial\xi<0$ near the axis,
a particle at $r=0$ is on an effective potential \textit{hill} as shown in the
$S=-0.5$ curve in Fig.\ \ref{fig:potential}(a) and will fall radially
\textit{out} of the flux tube, i.e., RUM. Since $\lambda=2B_{\phi}/B_{z}r$,
Eq.\ (\ref{S}) is identical to Eq.\ (\ref{equation of motion criteria}). All
magnetic flux tubes have uniform $B_{z}$ and $J_{z}$ near the axis so
particles with $S<0$ will always be on a potential hill and experience RUM.
Equation (\ref{S}) can be written in terms of experimental parameters as%
\begin{equation}
S=KIv_{z0}/q+0.5<0 \label{K}%
\end{equation}
where $K=\mu_{0}\pi mr^{2}/\Phi^{2}$ $\ $is positive, showing that RUM
requires $Iv_{z0}/q<0$, i.e., counter-current flow. Because $K\propto m$, ions
have a much lower threshold for RUM than electrons. 

%%%%%%%%%%%%%%%%%%%%%%%%%%%%%%%%%%%%%%%%%%
%-----FIGURE 3 (Particle trajectories)
%%%%%%%%%%%%%%%%%%%%%%%%%%%%%%%%%%%%%%%%%%
Figure \ref{fig:potential}(a) plots $V(\xi)$ given by Eq.\ (\ref{V})
for $S=-0.5$ and $1.5$, while Fig.\ \ref{fig:potential}(b) plots
trajectories calculated from direct numerical integration of $m\mathbf{\ddot
{r}=}q\mathbf{v\times}\left(  B_{\phi}\hat{\phi}+B_{z}\hat{z}\right)  $ for a
particle with $S=-0.5$ (i.e., $v_{z0}=-\omega_{c}/\lambda$) starting at the
down arrow and for a particle with $S=1.5$ (i.e., $v_{z0}=+\omega_{c}/\lambda
$) starting at the up arrow. Injection at $\lambda x=\lambda y=10^{-6}$ is
used so a particle does not start exactly at the top of a potential hill. To
approximate the weak field to the right of the flux loop sketched in
Fig.\ \ref{fig:coordinate}(b), an exponentially decaying $B_{z}$ in the
current-free external region is used in the numerical calculation. Figure
\ref{fig:potential}(b) shows that the $S=-0.5$ particle is ejected from the
flux tube (i.e., RUM), whereas the $S=1.5$ particle remains on the flux tube
axis (i.e., is confined).

Our experimental configuration \cite{Hansen04}, sketched in Fig.\ \ref{fig:coordinate}(b), 
involves top and bottom electrodes (respectively cathode and anode) mounted on the end dome 
of a large vacuum chamber (base pressure $\sim 10^{-7}$ mbar). 
The experimental sequence is: (i) slow ($\sim 10$ ms) electromagnets  behind the electrodes 
create an initial arched vacuum magnetic field, (ii) a fast ($\sim 1$ ms) gas valve injects
neutral gas from orifices in the electrodes, (iii) a $\sim 1$ kJ,  59 $\mu$F capacitor switched 
across the electrodes breaks down the neutral gas, (iv) a bright plasma loop appears. 
The 10--20 $\mu $s dynamical
evolution of this loop is imaged \cite{EPAPS} by a  fast digital framing camera.
Detailed  measurements in a similar experiment \cite{You} showed
that the bulk plasma in the flux loop is many orders of magnitude denser
than the injected pre-breakdown neutral gas and  results from fast MHD ingestion 
into the loop of  orifice-originating plasma \cite{Bellan05}. Figure 
\ref{fig:potential}(c) shows magnetic probe \cite{Romero} measurements of flux tube 
$B_{\phi }$ and $B_{z}$ as functions of distance $L$ from the flux tube axis in the
direction away from the electrode plane (data deconvolved as in
Ref.\ \cite{Burlaga}); the magnetic field amplitude decays rapidly
to the right (corresponding to weak field region in Fig.\ \ref{fig:coordinate}(b)).

\begin{figure}
\includegraphics{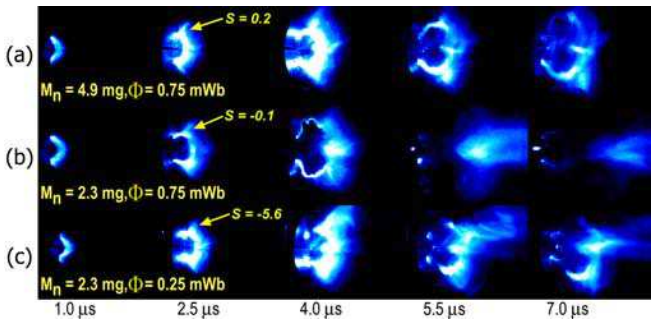}
\caption{(Color online)
Evolution of laboratory plasma loops at (a) high $M_{n}$, (b) low $M_{n}$, and
(c) low $\Phi$. Movies placed in Ref.\ \cite{EPAPS} show the
evolutions even more dramatically.}%
\label{fig:ar}%
\end{figure}

Figure \ref{fig:ar} shows the evolution of Ar plasma loops for different
injected gas mass $M_{n}$ and different flux $\Phi$ as a function of time
measured from breakdown. $M_{n}$ was determined using a thermocouple gauge and
has only a relative meaning because the plasma shot, being much shorter than
the gas puff time, uses only a fraction of $M_{n}$. Since the plasma has very
low impedance, the capacitor acts approximately as a current source. This is
consistent with the observation that $I$ and hence $\lambda/\omega_{c}$ are
essentially unaffected when $M_{n}$ is varied. However, the plasma 
velocity $v_{z0}$ is observed to be strongly dependent on $M_{n}$ with higher
$v_{z0}$ observed at smaller $M_{n}$.
%%%%%%%%%%%%%%%%%%%%%%%%%%%%%%%%%%%%%%%%%%
%-----FIGURE 4 (Argon pictures)
%%%%%%%%%%%%%%%%%%%%%%%%%%%%%%%%%%%%%%%%%%%

Figure \ref{fig:ar}(a) corresponds to $M_{n}=4.9$ mg, $\Phi=0.75$
mWb; Fig.\ \ref{fig:ar}(b) to $M_{n}=2.3$ mg, $\Phi=0.75$ mWb; and
Fig.\ \ref{fig:ar}(c) to $M_{n}=2.3$ mg, $\Phi=0.25$ mWb. In the first two
frames of Figs.\ \ref{fig:ar}(a-c) the plasma has a smooth arch shape; $I$ is
low at this stage and the plasma follows the half-torus profile of the initial
vacuum magnetic field spanning the electrodes. Then, as $I$ increases, the
plasma minor radius decreases due to self-pinching while the major radius
increases due to the hoop force \cite{Hansen01} associated with the poloidal
magnetic field produced by $I$. While this is happening, the loop undergoes
MHD kink instability and the projection of the writhed loop axis results in a
cusp-like dip at the apex \cite{Hansen01}. In the second frame (i.e., $2.5$
$\mu$s) of Figs.\ \ref{fig:ar}(a,b) a finger-like stream of plasma
emerges near the top (i.e., near cathode) of the loops. In Fig.\ \ref{fig:ar}%
(b), which corresponds to low $M_{n}$ and hence high $v_{z0}$, the stream
moves toward the ground plane near the cathode and leads to a major disruption
in $I$. As also seen in Fig.\ \ref{fig:ar}(b), this is followed by the
detachment of the loop from the electrodes and, for $t>4$ $\mu$s, formation of
a plasma jet propagating far to the right of the electrodes (see also
Fig.\ \ref{fig:jet}) into the weak field region (i.e., to the right in
Figs.\ \ref{fig:coordinate}(b) and \ref{fig:potential}(c)). A significant drop
in $I$ is observed during the detachment phase as well as an associated upward
voltage spike. From this time on, $I$ commutates to a new shorter path between
the electrodes, while the detached plasma jet propagates away from the
electrodes. When $\Phi$ is lowered as shown in Fig.\ \ref{fig:ar}(c),
two critical stages of the detachment are clearly seen in the 4.0--7.0 $\mu$s
frames, specifically the loop first detaches from the cathode and then from
the anode to form an intense plasma jet. Figures \ref{fig:ar}(a-c) also display $S$ estimated using
measured cathode region quantities in Eq.\ (\ref{K}) at 2.5 $\mu$s (i.e., just
before detachment) and indicate that the plasma jet development in
Figs.\ \ref{fig:ar}(b, c) is associated with having $S<0$.

Equation (\ref{K}) shows that only ions with $v_{z0}$ being large and negative
relative to $I$ can have $S<0$. Measurements (discussed below) indicate that
near the cathode ions with large negative $v_{z0}$ ($\sim40$--$60$ km/s) indeed exist. 
The slowing-down time ($>100~\mu$s) of these fast ions by the plasma (density $\sim 10^{20}$ m$^{-3}$) 
is much longer than the plasma duration, therefore collisions cannot affect their orbits. 
Since the ion contribution to electric current necessarily flows in the same direction as the current, 
ion drift motion associated with electric current cannot account for the observed
large negative $v_{z0}$. Furthermore, because the measured $\left\vert
v_{z0}\right\vert $ greatly exceeds the Ar$^{+}$ thermal speed $v_{T}\sim$2--5
km/s estimated using the spectroscopically determined $T_{i}\sim$1--10 eV,
neither can ion thermal motion account for the observed large negative
$v_{z0}.$ However, there does exist a mechanism capable of accelerating ions
to high velocities either parallel or anti-parallel to $I$. This mechanism
\cite{Bellan05,You} shows that axial gradients of $B_{\phi}^{2}$ provide an
MHD force $\sim-\partial B_{\phi}^{2}/\partial z$ that accelerates plasma
from regions of large $B_{\phi}^{2}$ to regions of small $B_{\phi}^{2}$, i.e.,
acceleration occurs from both foot-points of a flux loop towards the
apex if the flux loop minor radius is smaller at the foot-points than at the
apex (see detailed discussion in Ref.\thinspace\cite{Bellan05}). The resulting
velocity is $v_{z}\sim B_{\phi}/\sqrt{\mu_{0}m_{i}n_{i}}$, consistent with
higher ion axial velocity observed at smaller neutral gas injection pressures.
%%%%%%%%%%%%%%%%%%%%%%%%%%%%%%%%%%%%%%%%%%
%-----FIGURE 5 (spectrum)
\begin{figure}[t]
\includegraphics{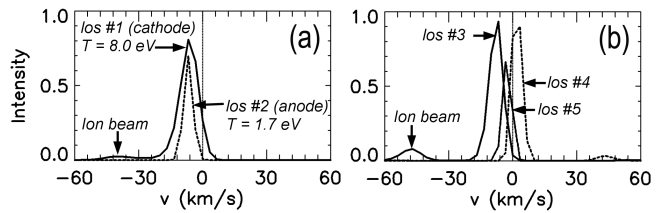}
\caption{Measured spectra
of an $\text{Ar}^{+}$ line (rest-frame wavelength $\lambda_{0}=434.806$ nm
shown by vertical lines). Velocity is $v=c(\lambda-\lambda_{0})/\lambda_{0}$
where $\lambda$ is measurement wavelength, $c$ is speed of light. Lines of
sight (los \#) are shown in Fig.\ \ref{fig:jet}. (a)
Spectra from the cathode (los \#1) and anode \ (los \#2) regions for $t$ =
0--0.5 $\mu$s. (b) Spectra from the kinetic jet region along los \#3, \#4, and
\#5 for $t$ = 2.5--5.5 $\mu$s, 7.0--9.0 $\mu$s, and 6.0--18.0 $\mu$s
respectively. }%
\label{fig:spectra}%
\end{figure}
%%%%%%%%%%%%%%%%%%%%%%%%%%%%%%%%%%%%%%%%%%

Ar$^{+}$ Doppler velocity measurements have been made using a 1 m
monochromator with a gated intensified CCD camera with fiber/lens coupling
system. The spectra displayed in Fig.\ \ref{fig:spectra} show velocity
components along lines of sight (los) indicated in Fig.\ \ref{fig:jet} by
\textquotedblleft los \#\textquotedblright. The los \#1 and \#2 spectra in
Fig.\ \ref{fig:spectra}(a) show that both cathode and anode emission lines are
blue-shifted, confirming suprathermal ion flow from \textit{both} cathode and
anode towards the apex as predicted by Ref.\thinspace\cite{Bellan05}. This
outflow is seen in camera images as 30--60 km/s bright fronts propagating away
from both electrodes along the flux loop axis towards its apex \cite{EPAPS}.
Figure \ref{fig:spectra}(a) shows a large ion velocity component
($\sim40$ km/s ion beam for los \#1) moving away from the cathode while
Fig.\ \ref{fig:spectra}(b) shows spectra measured in the kinetic jet
region and indicates a $\sim50$ km/s ion beam for los \#3. Because, as seen in
Fig.\ \ref{fig:jet}, los \#1 and los \#3 make different angles
relative to the respective flow directions being measured, ion beam velocities
between different los \#'s cannot be quantitatively compared, i.e., the $50$
km/s los \#3 ion beam in Fig.\ \ref{fig:spectra}(b) cannot be
interpreted as a 10 km/s acceleration of the 40 km/s los \#1 ion beam in
Fig.\ \ref{fig:spectra}(a).

%%%%%%%%%%%%%%%%%%%%%%%%%%%%%%%%%%%%%%%%%%
%-----FIGURE 6 (S-Parameter)
\begin{figure}
\includegraphics{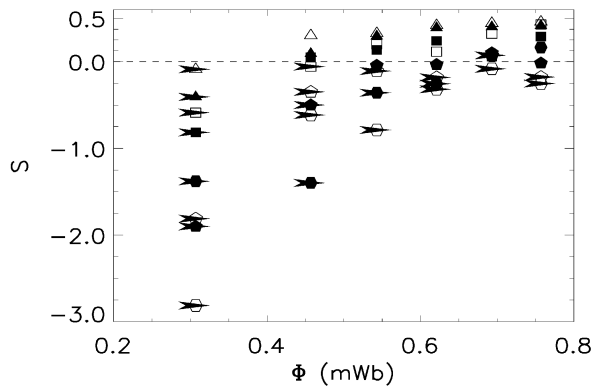}
\caption{Measured $S$ for
various plasma configurations. Filled/open symbols represent high/low $M_{n}$
respectively, number of sides in symbols represent capacitor charging voltage
in kV, and arrow heads indicate plasma shots where kinetic jets are observed.
Existence of kinetic jets has an excellent correlation with $S$ being
negative.}%
\label{fig:S}%
\end{figure}
%%%%%%%%%%%%%%%%%%%%%%%%%%%%%%%%%%%%%%%%%%
Figure \ref{fig:S} shows the results of a parameter scan of $I,\Phi,$
and $v_{z0}$ performed to determine the $S$ dependence of the instability
onset. $v_{z0}$ is determined from plasma front motion in the camera images
\cite{EPAPS}. The $S$ values in Fig.\ \ref{fig:S} were calculated
using cathode region parameters in Eq.\ (\ref{K}) for a large number of argon
plasma loops ($r\simeq8$ mm); $S=0.5$ is the upper bound for the observed
negative $Iv_{z0}$. Kinetic jet instability, shown by arrow-heads in
Fig.\ \ref{fig:S}, occurs only when $S<0$ indicating excellent
agreement with the RUM onset prediction.

The plasma loops used for Fig.\ \ref{fig:S} have already undergone MHD
kink instability \cite{Hsu} since all have $\left\vert \lambda\right\vert
>4\pi/L\simeq60$ m$^{-1}$, where $L\simeq$ 0.2 m is the loop length. The loops
produce kinetic jets only when $S<0$ showing that RUM is a kinetic, rather
than MHD, instability. The kinetic nature is also evident from the high
velocity beams in Fig.\ \ref{fig:spectra} and from the kinetic jet
appearing in the weak field (non-MHD) region as 
sketched in Fig.\ \ref{fig:coordinate}(b).

The RUM model explains why counter-injected neutral beams in tokamaks have
severe orbit losses compared to co-injected neutral beams
\cite{Mikkelsen,Egedal,McClements}. In particular, Fig.\ 10 of
Ref.\ \cite{Mikkelsen} showed that an 80 keV counter-injected
deuterium beam has severe orbit losses in a $B=0.3$ T tokamak having safety
factor $q=1.25$ and major radius $R\simeq1$ m. Since $\lambda
\simeq2/qR\simeq1.6$ m$^{-1},$ $\omega_{cD}=\allowbreak1.\,\allowbreak
4\times10^{7}$ s$^{-1},$ and the injection velocity is $v_{inj}\ =\allowbreak
2.\,\allowbreak8\times10^{6}$ m/s, it is seen that $S_{counter} \simeq
 0.5-\lambda v_{inj}/\omega_{cD}\ \approx0.2$ whereas $S_{co}=0.5+\lambda
v_{inj}/\omega_{cD}=0.8$; so, counter-injected ions have much larger orbits
(i.e., broader valley-type effective potential as in Fig.\ \ref{fig:potential}%
(a)) than co-injected ions. While coronal loops are unlikely
to have $S<0$ due to their small $\lambda$ ($\sim10^{-8}$ m$^{-1}$)
\cite{Burnette}, jets associated with canceling magnetic features
\cite{Chae,Liu,Harrison} and coronal streamers \cite{Li} emanating near
magnetic neutral lines are both produced in extremely low magnetic field
regions where $S<0$ could occur and RUM\ may be operative.

In summary, an instability has been demonstrated where ions are
\textit{magnetically ejected} from a flux tube. Ejection occurs when ions move
opposite to the current with a sufficiently large axial velocity. We thank
A.\ H.\ Boozer for pointing out a relationship between RUM and neutral beam
counter-injection and D.\ Felt for technical assistance. Supported by US DOE
and by NSF.

%%%%%%%%%%%%%%%%%%%%%%%%%%%%%%%%%%%%%%%%%%


\begin{thebibliography}{99}  
\bibitem {Tandberg}E. Tandberg-Hanssen, \textit{The Nature of Solar
Prominences} (Kluwer Academic, Dordrecht, 1995).

\bibitem {Bellan_book2000}P. M. Bellan, \textit{Spheromaks} (Imperial College
Press, London, 2000).

\bibitem {Tokamak}J. Sheffield, Rev. Mod. Phys. \textbf{66,} 1015 (1994).

\bibitem {Burlaga}L. F. Burlaga, J. Geophys. Res. \textbf{93,} 7217 (1988).

\bibitem {Mikkelsen}D. R. Mikkelsen et al., Phys. Plasmas \textbf{4,} 3667 (1997).

\bibitem {Egedal}J. Egedal et al., Phys. Plasmas \textbf{10}, 2372 (2003).

\bibitem {McClements}K. G. McClements and A. Thyagaraja, Phys. Plasmas
\textbf{13,} 042503 (2006).

\bibitem {Chae}J. Chae, Astrophys. J. \textbf{584,} 1084 (2003).

\bibitem {Liu}Y. Liu and H. Kurokawa, Astrophys. J., \textbf{610, }1136 (2004).

\bibitem {Harrison}R. A. Harrison, P. Bryans, and R. Bingham, Astron.
Astrophys. \textbf{379,} 324 (2001).

\bibitem {Li}J. Li et al., Astrophys. J. \textbf{506,} 431 (1998).

\bibitem {Hansen04}J. F. Hansen, S. K. P. Tripathi, and P. M. Bellan, Phys.
Plasmas \textbf{11,} 3177 (2004).

\bibitem {EPAPS}See EPAPS Document No. \hspace{0.3in} for movies and camera
$v_{z0}$ measurement example.

\bibitem {You}S. You, G. S. Yun, and P. M. Bellan, Phys. Rev. Lett.
\textbf{95,} 45002 (2005).

\bibitem {Bellan05}P. M. Bellan, Phys. Plasmas \textbf{10,} 1999 (2003).

\bibitem {Romero}C. A. Romero-Talamas, P. M. Bellan, and S. C. Hsu, Rev. Sci.
Instrum. \textbf{75}, 2664 (2004).

\bibitem {Hansen01}J. F. Hansen and P. M. Bellan, Astrophys. J. \textbf{563,}
L183 (2001).

\bibitem {Hsu}S. C. Hsu and P. M. Bellan, Phys. Rev. Lett. \textbf{90}, 215002 (2003).

\bibitem {Burnette}A. B. Burnette and R. C. Canfield, Astrophys. J.
\textbf{606,} 565 (2004).
\end{thebibliography}
\end{document}